\documentclass[aps,prd,floats,preprint,tightenlines,nofootinbib]{revtex4}
\newcommand{\tr}{{\rm Tr}}
\newcommand{\sfrac}[2]{{\textstyle\frac{#1}{#2}}}

\begin{document}
\preprint{ \hbox{hep-ph/0602154} } \vspace*{3cm}

\title{Disguising the Oblique Parameters}
\author{ Christophe Grojean$^{a,b}$, Witold Skiba$^c$, and John Terning$^d$}
\affiliation{  \small \sl  
		$^a$~Service de Physique Th\'eorique, CEA Saclay,
     			F-91191 Gif-sur-Yvette, France \\
     $^b$~CERN Physics Department, Theory Division, CH-1211 Geneva 23, Switzerland\\
               $^c$~Department of Physics, Yale University, New Haven, CT  06520, USA \\
               $^d$~Department of Physics, University of California, Davis, CA 95616, USA
                                \vspace{2.5cm}
            }

\begin{abstract}
We point out a set of operator identities that relate the operators corresponding to the oblique
corrections to operators that modify fermion couplings to the gauge bosons as well as operators
that modify triple gauge boson couplings. Such identities are simple consequences of the 
equations of motion.  Therefore the contributions from new physics to the oblique parameters
can be disguised as modifications of triple gauge boson couplings provided the fermion 
couplings to the gauge bosons are suitably modified by higher dimensional operators.
Since the experimental constraints on triple gauge boson couplings are much weaker than the
constraints on the oblique parameters this observation allows extra room for model building. 
We derive operator relations in effective theories of the Standard Model 
with the electroweak symmetry either linearly or nonlinearly realized and discuss applications of our results.
\end{abstract}

\maketitle

\newpage

\section{Introduction}
The consequences of new heavy particles
in extensions of the Standard Model (SM) can be accounted for at low energies in terms of new effective operators.
The wealth of data collected by the LEP, SLD, and many other experiments severely constrains
new operators involving the electroweak sector.  The most widely 
used operators for constraining new models are those that modify the gauge boson two-point functions,
which are often referred to as the oblique~\cite{oblique,STU}, or universal~\cite{LEP12}, corrections.
Among several parameterizations of the gauge boson two-point functions, the $S$, $T$, $U$
parameters ~\cite{STU} are the best known. The constraints on the
$S$ and $T$ parameters  are some of the most stringent among  operators of
the same dimension~\cite{Han:2004az}. Tight constraints on the oblique parameters pose
challenges for many extensions of the SM.

We want to point out that, since a shift in the couplings between fermions and gauge bosons can be absorbed as oblique corrections, the observable effects of the oblique parameters can be, 
to a large degree, removed if fermion couplings to gauge bosons are also modified (related observations have been made in Refs.~\cite{Previous}). 
That is to say non-oblique corrections can obscure  the oblique corrections
and new physics generating non-oblique corrections can mask other effects that produce oblique corrections.   We consider a complete set of higher-dimensional operators added to the SM Lagrangian and we show that
the lowest order equations of motion imply relations between higher-dimensional operators corresponding to
the oblique operators, operators modifying the fermion couplings, and operators modifying the
triple gauge boson couplings.  It seems to us that although this fact is known its usefulness
is not widely appreciated. The triple gauge boson couplings were measured by the LEP2 experiments,
but the statistics are much more limited compared to the $Z$-pole data. Therefore suitable modifications
of couplings of fermions to the gauge bosons can render the constraints on many models much
milder than one would have anticipated if only the oblique parameters were considered.
To turn things around, this implies that improving the electroweak constraints will certainly
require a better knowledge of triple gauge boson couplings. A task that will have to wait for
the linear collider.

One might think that the electroweak precision measurements will become moot as we enter
the LHC era. This is not likely to happen even after the discovery of new particles. At best, we will gain
a partial knowledge about the spectrum of new particles at the LHC. We will learn very little
about the couplings of the newly discovered states. Electroweak precision tests will continue
guiding us towards the right theory.

In this paper we will discuss the operator relations in the frameworks of effective theories of the SM both 
with  linearly and nonlinearly realized electroweak symmetry. In the case of linearly realized
symmetry only $S$ and $T$ are relevant as they correspond to dimension 6 operators, while 
$U$ corresponds to a dimension 8 operator. The equations of motion for the $SU(2)\times U(1)$ 
gauge fields yield two independent relations involving $S$ and $T$. In the case of nonlinearly
realized symmetry the same equations of motion yield three independent relations involving
$S$, $T$, and $U$. The main point of this article can be understood by glancing at 
Eqs.~(\ref{eq:lf1}) and  (\ref{eq:lf2}) in the linear realization, 
and Eqs.~(\ref{eq:nf1})--(\ref{eq:nf3}) in the nonlinear realization.

In the next two sections we discuss in turn the linear and nonlinear cases. In Sec.~\ref{sec:Applications}
we give a toy example and discuss applications to extra dimensional scenarios.
We summarize the lowest dimensional  chiral Lagrangian for the electroweak
theory in Appendix~\ref{app:Lagrangian}.

\section{Linear case}
\label{sec:Linear}

We now turn to an effective theory containing the SM fields with one Higgs doublet. 
The most important higher-dimensional operators have dimension 6. A Majorana mass term
for the left-handed neutrinos has dimension 5, but we are only interested in flavor-preserving
operators since the oblique parameters are flavor universal. It is straightforward to enumerate
all operators of dimension 6, see Ref.~\cite{Buchmuller}. Integration by parts and equations
of motion are used extensively to avoid redundancy among operators. 
What we now show is that using the equations of motion, we relate particular linear combinations involving the oblique parameters and other operators to peculiar redundant operators that 
only affect the triple gauge boson couplings.

We will use the notation of Ref.~\cite{Buchmuller}.  We will need only a small subset of operators
in Ref.~\cite{Buchmuller} 
 \begin{eqnarray} 
    O_{W\!B}=(h^\dagger \sigma^a h) W^a_{\mu \nu} B^{\mu \nu},  && 
    O_h = | h^\dagger D_\mu h|^2, \label{eq:owbh} \\
   O_{hl}^s = i (h^\dagger D^\mu h)(\overline{l} \gamma_\mu l) + {\rm h.c.}, &&
   O_{hl}^t = i (h^\dagger \sigma^a D^\mu h)(\overline{l} \gamma_\mu \sigma^a l)+ {\rm h.c.},
     \label{eq:ohl} \\ 
   O_{hq}^s = i (h^\dagger D^\mu h)(\overline{q} \gamma_\mu q)+ {\rm h.c.}, &&
   O_{hq}^t = i (h^\dagger \sigma^a D^\mu h)(\overline{q} \gamma_\mu \sigma^a q)+ {\rm h.c.},
    \label{eq:ohq} \\ 
      O_{hu} = i (h^\dagger D^\mu h)(\overline{u} \gamma_\mu u)+ {\rm h.c.}, &&
   O_{hd} = i (h^\dagger D^\mu h)(\overline{d} \gamma_\mu d)+ {\rm h.c.},
           \label{eq:Ohud}  \\ 
   O_{he} = i (h^\dagger D^\mu h)(\overline{e} \gamma_\mu e)+ {\rm h.c.}\,. &&
     \label{eq:ohe}
\end{eqnarray}
where $W^a_{\mu \nu}$ is the $SU(2)$ field strength, $B_{\mu \nu}$ the hypercharge field strength
and $h$ represents the Higgs doublet. The left-handed fermions are denoted $q$ and $l$, while
the right-handed ones $u$, $d$, and $e$. The family indices are implicitly summed over all three families.
$O_{W\!B}$ corresponds to the $S$ parameter and $O_h$ to $T$. The remaining operators on our list
alter fermion couplings to the $B$ and $W$ gauge bosons.

The lowest-order Lagrangian is
\begin{equation}
  {\mathcal L}={\mathcal L}_{gauge-fermion} + (D^\mu h)^\dagger (D_\mu h) - V(h)
  \label{eq:lLagrangian}
\end{equation}
and the corresponding equations of motion for the gauge bosons are
\begin{eqnarray}
   \partial^\mu B_{\mu \nu}  + i \frac{g'}{2} (h^\dagger D_\nu h - D_\nu h^\dagger h) + 
         g' \sum_f Y_f \overline{f} \gamma_\nu  f & = & 0, \label{eq:lBeom}  \\
   D^\mu W_{\mu \nu}^a + i \frac{g}{2} (h^\dagger  \sigma^a D_\nu h - D_\nu h^\dagger \sigma^a h) + 
       \frac{g}{2} \sum_f \overline{f}_L \gamma_\nu \sigma^a f_L & = & 0, \label{eq:lWeom}
 \end{eqnarray}
where $Y_f$ is the hypercharge of fermion $f$.

Multiplying Eq.~(\ref{eq:lBeom}) by $(i h^\dagger D^\nu h +  {\rm h.c.})$ and Eq.~(\ref{eq:lWeom}) 
by $(i h^\dagger \sigma^a D^\nu h +  {\rm h.c.})$ we obtain
\begin{eqnarray}
 \nonumber
 2 g' O_h - \frac{g}{2} O_{W\!B} + g'  O_{hf}^Y&=& 2 i B_{\mu \nu} D^\mu h^\dagger D^\nu h 
 -  g' h^\dagger h \, D^\mu h^\dagger D_\mu h 
 \\
 &&
+\frac{g'}{2}  h^\dagger h  (B_{\mu \nu})^2
 - \frac{g'}{2} h^\dagger h \left( h^\dagger D^2 h +  (D^2 h^\dagger) h \right),
         \label{eq:l1} \\
\nonumber
 -g' O_{W\!B} +g (O_{hl}^t+O_{hq}^t) 
    &= &
 4 i W^a_{\mu \nu} D^\mu h^\dagger \sigma^a D^\nu h
  - 6 g  h^\dagger h\, D^\mu h^\dagger D_\mu h  \\
&&
 +g  h^\dagger h  (W_{\mu \nu}^a)^2 
 - g h^\dagger h \left( h^\dagger D^2 h +  (D^2 h^\dagger) h \right),
         \label{eq:l2} 
\end{eqnarray}
where $O_{hf}^Y=\sum_f Y_f O_{hf}^s=\frac{1}{6} O_{hq}^s-\frac{1}{2} O_{hl}^s +\frac{2}{3} O_{hu} 
                -\frac{1}{3} O_{hd} - O_{he}$.  Note that of the four terms on the right-hand sides of Eqs.~(\ref{eq:l1}) and (\ref{eq:l2})
the first terms are observable as they modify gauge boson self couplings, while the second and third terms are not currently observable and the fourth term gives a contribution proportional to the fermion masses that we neglect. The second and third  and part of the fourth terms renormalize the lowest-order Lagrangian  in  
Eq.~(\ref{eq:lLagrangian}) when $h^\dagger h$ is substituted by its vacuum expectation value, $v$.

To make the operator relations more transparent let us redefine the normalization of operators
$O_{W\!B}$ and $O_{h}$ as follows
\begin{equation}
    O_{S}= \frac{\alpha}{4 s c v^2} O_{W\!B},  \ \ \  O_{T}=  -\frac{2\alpha}{v^2} O_h,
\end{equation}
where $s$ and $c$ are the sine and cosine of the weak mixing angle, $\alpha$ is the fine structure
constant, and $v\approx 250~{\rm GeV}$. 
The rescaled operators $ O_{S}$ and $ O_{T}$ are defined 
such that their coefficients are, respectively, the $S$ and $T$ parameters. That is the Lagrangian including the operators  $\mathcal{L}=a_S O_S + a_T O_T$ gives
a contribution to $S$ and $T$ equal to $S=a_S$ and $T=a_T$. 
Neglecting the unobservable terms in Eqs.~(\ref{eq:l1}) and (\ref{eq:l2}) we get
\begin{eqnarray}
  - \frac{2 g s c v^2}{\alpha}  O_{S} - \frac{g' v^2}{\alpha}  O_{T} + g'  O_{hf}^Y
         &=& 2 i B_{\mu \nu} D^\mu h^\dagger D^\nu h, 
 \label{eq:lf1}\\
   - \frac{4 g'  s c v^2}{\alpha}  O_{S}  +g (O_{hl}^t+O_{hq}^t)
         &=& 4 i W^a_{\mu \nu} D^\mu h^\dagger \sigma^a D^\nu h. 
\label{eq:lf2}
\end{eqnarray}
The relations~(\ref{eq:lf1})--(\ref{eq:lf2}) can be understood as an equivalence between oblique corrections and shifts in the fermion couplings to gauge bosons up to modification of gauge boson self couplings. 
It is straightforward to convert the operators on the right-hand sides to the well-known parameterization 
of general triple gauge boson couplings in Ref.~\cite{Hagiwara:1986vm} and ensuing to contributions to
the $e^+ e^- \rightarrow W^+ W^-$ scattering cross sections.

Alternatively, the same operator relations can be obtained by field redefinitions~\cite{field:red}.
Let us consider the Lagrangian:
\begin{eqnarray}
\mathcal{L}= 
 -\sfrac{1}{4} W^a_{\mu\nu} W^{a\,\mu\nu} -\sfrac{1}{4} B_{\mu\nu} B^{\mu\nu} 
  + D_\mu h^\dagger D^\mu h + \sum_f i \bar{f}  D \!\!\!\! / f
  + \frac{g^\prime \epsilon_1}{\Lambda^2} O^Y_{hf} 
   + \frac{g \epsilon_2}{2\Lambda^2} (O^t_{hl} + O^t_{hq}).
\end{eqnarray}
Using the field redefinitions
\begin{eqnarray}  
\hat{B}_\mu & = & B_\mu 
      + \frac{\epsilon_1}{\Lambda^2} i (h^\dagger D_\mu h -  D_\mu h^\dagger h), 
\label{eq:fred1}
      \\
\hat{W}^a_\mu &= & W^a_\mu 
	+ \frac{\epsilon_2}{\Lambda^2}
         i (h^\dagger \sigma^a D_\mu h -   D_\mu h^\dagger \sigma^a h), 
\label{eq:fred2}
\end{eqnarray}
the Lagrangian now reads 
\begin{eqnarray}
\mathcal{L}&= &
         -\sfrac{1}{4} \hat{W}^a_{\mu\nu} \hat{W}^{a\,\mu\nu}
          -\sfrac{1}{4} \hat{B}_{\mu\nu} \hat{B}^{\mu\nu} 
          + \hat{D}_\mu h^\dagger \hat{D} h + \sum_f  i \bar{f}  \hat{D} \!\!\!\! / f
          - \frac{2g' \epsilon_1}{\Lambda^2} O_{h} 
         + \frac{g \epsilon_1+g^\prime \epsilon_2}{\Lambda^2}   O_{WB}\nonumber\\
&&     +\ldots
\end{eqnarray}
where the hatted quantities denote field strengths and covariant derivaties associated to the 
$\hat{W}_\mu^a$ and $\hat{B}_\mu$ gauge fields and the $\ldots$ stands for the unobservable
terms  as well as for the operators modifying the triple gauge boson self couplings appearing
on the right hand side of Eqs.~(\ref{eq:l1})--(\ref{eq:l2}).

Let us comment on the result of Ref.~\cite{Han:2004az} where bounds on arbitrary linear combinations
of the operators in Eqs.~(\ref{eq:owbh})--(\ref{eq:ohe}) were obtained. It is easy to check that the linear
combinations of operators on the left-hand sides of Eqs.~(\ref{eq:lf1}) and (\ref{eq:lf2}) are relatively
weakly constrained. Suppose these linear combinations of operators are added to the SM
Lagrangian one at a time with a coefficient $\frac{1}{\Lambda^2}$. The $90\%$ confidence level
bounds are  $\Lambda> 650~{\rm GeV}$ for the coefficient multiplying the operators in Eq.~(\ref{eq:lf1}), 
and $\Lambda> 1.2~{\rm TeV}$ for the ones in Eq.~(\ref{eq:lf2}). The only source of these
bounds is the data on the  $e^+ e^-\rightarrow W^+ W^-$ cross section. 
If the data on the $e^+ e^-\rightarrow W^+ W^-$ scattering was not used in
Ref.~\cite{Han:2004az}  these linear combinations of operators would not be bounded at all.

The analysis of gauge boson self energies was recently extended in Ref.~\cite{LEP12} to include
higher-derivative terms. The relevant higher-derivative terms are governed by the operators
$O_{BB}=\frac{1}{2} \partial_\alpha B_{\beta \gamma} \partial^\alpha B^{\beta \gamma}$ and 
$O_{WW}=\frac{1}{2} D_\alpha W^a_{\beta \gamma} D^\alpha W^{a \beta \gamma}$.
These operators are not listed as independent operators in Ref~\cite{Buchmuller} because, using  Bianchi identities like $\partial_\alpha B_{\mu\nu} + \partial_\mu B_{\nu\alpha}+
\partial_\nu B_{\alpha\mu}=0$,
they are equal to $( \partial^\mu B_{\mu \nu} )^2$ and $( D^\mu W_{\mu \nu}^a)^2$, respectively,
so they can be expressed as the squares of Eqs.~(\ref{eq:lBeom}) and (\ref{eq:lWeom}),
see also Ref.~\cite{Strumia:1999jm}. Therefore, $O_{BB}$ and $O_{WW}$ are equivalent to linear 
combinations of the operators in Eqs.~(\ref{eq:owbh})--(\ref{eq:ohe}) as well as four-fermi operators. 
We are not aware of any further identities that would relate $O_{BB}$ or $O_{WW}$ to
triple gauge boson couplings in analogy with the $S$ and $T$ parameters.
 
\section{Nonlinear case}
 \label{sec:Nonlinear}
The scalar sector of the SM without the physical Higgs boson is conveniently described by a $\Sigma$ field
\begin{equation}
  \Sigma=\exp{\left(\frac{i \pi^a \sigma^a}{v}\right)},
\end{equation}
where $\sigma^a$ are the Pauli matrices. $\Sigma$ transforms linearly under the $SU(2)_L\times SU(2)_R$
as $\Sigma \rightarrow L \Sigma R^\dagger$. The hypercharge is embedded in $SU(2)_R$, so that 
$D_\mu \Sigma = \partial_\mu \Sigma - i g W_\mu \Sigma + \frac{i}{2} g' B_\mu \Sigma \sigma^3$,
where $W_{\mu} =W^a_{\mu} \frac{\sigma^a}{2}$. The higher dimensional operators
of interest to us fall into two classes. First, operators containing the gauge bosons and
the $\Sigma$ fields~\cite{oblique,AppelquistBernard,Longhitano,AppelquistWu}.
These include the operators corresponding to the oblique  parameters
as well as operators that modify higher-point gauge couplings. Second, operators containing 
two fermions, gauge fields, and the $\Sigma$ field~\cite{ABCH, Bagan:1998vu}. 
These operators modify the couplings of fermions
to the gauge fields. Appendix~\ref{app:Lagrangian} contains a list of all such operators.
To distinguish from the linear case operators are denoted as ${\mathcal L}$ instead of $O$.

It is more transparent to trade the $\Sigma$ field for the following combinations
\begin{eqnarray}
  V_\mu \equiv (D_\mu \Sigma) \Sigma^\dagger, & \ \ \ & T\equiv \Sigma \sigma^3 \Sigma^\dagger, \\
  \hat{V}_\mu \equiv (D_\mu \Sigma^\dagger) \Sigma, & \ \ \ &  \hat{T}\equiv \sigma^3.
\end{eqnarray}
In terms of these objects, the lowest-order Lagrangian is
\begin{equation}
  {\mathcal L}={\mathcal L}_{gauge-fermion}- \frac{v^2}{4} \tr(V^\mu V_\mu). 
      \label{eq:nLagrangian}
\end{equation}
In addition to $ \tr(V^\mu V_\mu)$ there exists another operator of dimension 2,
denoted ${\mathcal L}_0$ in appendix~\ref{app:Lagrangian},
but since ${\mathcal L}_0$  violates the custodial symmetry it is assumed that its coefficient is small.
The  corresponding equations of motion for $B_\nu$ and $W^a_\nu$ are
\begin{eqnarray}
   \partial^\mu B_{\mu \nu}  - i \frac{v^2 g'}{4} \tr(V_\nu T) + 
         g' \sum_f Y_f \overline{f} \gamma_\nu  f & = & 0, \label{eq:nBeom}  \\
   D^\mu W_{\mu \nu}^a + i \frac{v^2 g}{4} \tr(V_\nu \sigma^a) + 
       \frac{g}{2} \sum_f \overline{f}_L \gamma_\nu \sigma^a f_L & = & 0. \label{eq:nWeom}
 \end{eqnarray}
 
We multiply Eq.~(\ref{eq:nBeom}) by $i g' \tr(T V^\nu)$ and  Eq.~(\ref{eq:nWeom}) by
$i g \tr(V^\nu \sigma^a)$ as well as by $i g \tr(T V^\nu T \sigma^a)$. After a bit of algebra we obtain
\begin{eqnarray}
                 g'^2  {\mathcal L}_0 -  {\mathcal L}_1  + 
                g'^2 {\mathcal L}^Y_f
                  &=&  -{\mathcal L}_2 - \frac{g'^2}{2} B^{\mu \nu} B_{\mu \nu} , \label{eq:n1} \\
   {\mathcal L}_1 +g^2  ({\mathcal L}_q^1 + {\mathcal L}_l^1)  &=& {\mathcal L}_3 + 
                 g^2 \tr (W^{\mu \nu} W_{\mu \nu}) + \frac{g^2 v^2}{2} \tr(V^\mu V_\mu) ,   \label{eq:n2} \\ 
    -2 g^2  {\mathcal L}_0 +   {\mathcal L}_1 - 4 {\mathcal L}_8 +
                g^2  ({\mathcal L}_q^3 + {\mathcal L}_l^3)   
                &=&   {\mathcal L}_3 - 4  {\mathcal L}_9 - \frac{g^2 v^2}{2} \tr(V^\mu V_\mu), \label{eq:n3}
\end{eqnarray}
where $ {\mathcal L}_f^Y=\frac{1}{6}({\mathcal L}_q^2-{\mathcal L}_q^5) 
                     - \frac{1}{2}({\mathcal L}_l^2-{\mathcal L}_l^5)
                    -(  {\mathcal L}_q^4+{\mathcal L}_q^6+{\mathcal L}_l^4+{\mathcal L}_l^6)$ is the 
product of the fermion hypercharge current and $i \tr(T V^\nu)$. The operators on the right-hand sides 
of Eqs.~(\ref{eq:n1})--(\ref{eq:n3}) that are not abbreviated as ${\mathcal L}_i$ are terms in the
Lagrangian in Eq.~(\ref{eq:nLagrangian}) and therefore are not observable. 
Eqs.~(\ref{eq:n1})--(\ref{eq:n3})
also imply that the operators ${\mathcal L}_{2,3,9}$ are redundant, an observation already made in 
Ref.~\cite{Nyffeler:1999ap}. What is important for us is that ${\mathcal L}_{2,3,9}$ alter only the 
gauge boson self couplings. 

In analogy with the linear case discussed in the previous section, we define $ {\mathcal L}_{S}$ to be
the operator whose coefficient is the $S$ parameter, and so on for $T$ and $U$.
\begin{equation}
    {\mathcal L}_{S} = -\frac{1}{16 \pi} {\mathcal L}_1, \ \ \  {\mathcal L}_{T} =\frac{\alpha}{2} {\mathcal L}_0, \ \ \ 
   {\mathcal L}_{U} =-\frac{1}{16 \pi} {\mathcal L}_8.
\end{equation}
Neglecting the unobservable terms in Eqs.~(\ref{eq:n1})--(\ref{eq:n3}) we get
\begin{eqnarray}
   16 \pi   {\mathcal L}_{S}  + \frac{8 \pi}{c^2}  {\mathcal L}_{T}   +  g'^2 {\mathcal L}^Y_f 
          &=&  - {\mathcal L}_2, \label{eq:nf1} \\
    - 16 \pi   {\mathcal L}_{S} + g^2  ({\mathcal L}_q^1 + {\mathcal L}_l^1)  &=& {\mathcal L}_3,  \label{eq:nf2} \\
    - 16 \pi   {\mathcal L}_{S} - \frac{16 \pi}{s^2}  {\mathcal L}_{T}  + 64 \pi {\mathcal L}_{U}  + 
          g^2  ({\mathcal L}_q^3 + {\mathcal L}_l^3)
           &=&   {\mathcal L}_3 - 4  {\mathcal L}_9. \label{eq:nf3} 
\end{eqnarray}
As in the linear case, these relations can be obtained by field redefinitions. 
Indeed, the Lagrangian:
\begin{eqnarray}
\mathcal{L}&=& 
 -\sfrac{1}{4} W^a_{\mu\nu} W^{a\,\mu\nu} -\sfrac{1}{4} B_{\mu\nu} B^{\mu\nu} 
  -\frac{v^2}{4} \mathrm{Tr} (V^\mu V_\mu) + \sum_f i \bar{f}  D \!\!\!\! / f 
\nonumber \\
 && + g'^2 \epsilon_1 \mathcal{L}^Y_{f} 
  + g^2 \epsilon_2 (\mathcal{L}^1_q+\mathcal{L}^1_l)
  + g^2 \epsilon_3 (\mathcal{L}^3_q+\mathcal{L}^3_l),
\end{eqnarray}
can be brought back to its more canonical from
\begin{eqnarray}
\mathcal{L}&= &
               -\sfrac{1}{4} \hat{W}^a_{\mu\nu} \hat{W}^{a\,\mu\nu}
                -\sfrac{1}{4} \hat{B}_{\mu\nu} \hat{B}^{\mu\nu} 
                 -\frac{v^2}{4} \mathrm{Tr} (\hat{V}^\mu \hat{V}_\mu) + \sum_f  i \bar{f}  \hat{D} \!\!\!\! / f 
\nonumber\\
         &&+ (-g'^2 \epsilon_1 + 2 g^2 \epsilon_3 ) \mathcal{L}_{0}
                +  (\epsilon_1-\epsilon_2-\epsilon_3 ) \mathcal{L}_{1}
               +  4  \epsilon_3 \mathcal{L}_{8}
                 +\ldots
\end{eqnarray}
by the field redefinitions:
\begin{eqnarray}  
   \hat{B}_\mu & = & B_\mu + \epsilon_1 i g' \tr(T V_\mu), \\
  \hat{W}^a_\mu & = &W^a_\mu + \epsilon_2 i g \tr(V_\mu \sigma^a)
  + \epsilon_3 i g \tr(T V_\mu T \sigma^a).
\end{eqnarray}

\section{Applications}
\label{sec:Applications}

We want to illustrate how one might use our results in a toy example, and comment
how some of the results have already been incorporated into models with extra dimensions.

Let us start with a toy example. An electroweak triplet scalar breaks the custodial symmetry and 
therefore generates a contribution to the $T$ parameter. Hence one expects stringent constraints
on the couplings and mass of such a field. Suppose we introduce a complex triplet scalar  
with the hypercharge $-1$ that couples to the Higgs doublet as follows
\begin{equation}
   V=M^2_\phi \phi^{a*} \phi^a + \mu\left( \phi^a \tilde{h}^\dagger \sigma^a h +  
                    \phi^{a*} h^\dagger \sigma^a \tilde{h} \right),
\end{equation}
where $\tilde{h}=i \sigma^2 h^*$.
Assuming that $M_\phi$ is large we integrate $\phi$ out and keep the interesting part 
of the effective action 
\begin{equation}
  {\mathcal L}_\phi= \frac{4 \mu^2}{M^4_\phi} O_h , \label{eq:scalar}
\end{equation}
which corresponds to a negative correction to the $T$ parameter.
If $\phi$ were the only source of new physics, we would then get a 90\% confidence level limit
\begin{equation}
    \frac{\mu^2}{M^4_\phi} < \frac{1}{(14~\mathrm{TeV})^2}. \label{eq:tripletbound}
\end{equation}
If we want to relax this bound, we could add new particles that would give a positive contribution to $T$  to compensate. One could, for instance introduce a hypercharge $0$ weak triplet.
Alternatively, we can use the results presented in Section~\ref{sec:Linear} and look for new physics that will modify the fermion-gauge boson coupling in order not to cancel the coefficient of the $O_h$ operator but in order to generate a particular linear combination of higher dimensional operators that is poorly constrained.

Combining Eqs.~(\ref{eq:l1}) and (\ref{eq:l2}) gives a relation that only involves the operator  $O_h$ and
does not involve $O_{W\!B}$
\begin{equation}
    2 g'^2 O_h + g'^2 O_{hf}^Y - \frac{g^2}{2} (O_{hl}^t+ O_{hq}^t)=O_{3V}, \label{eq:eliminateT}
\end{equation}
where $O_{3V}$ indicates a linear combination of operators on the right-hand sides of 
Eqs.~(\ref{eq:l1}) and (\ref{eq:l2}) that modify gauge boson self couplings. This particular linear combination
of operators is much less constrained by the precision electroweak data than the coefficient of $O_h$.
Let us make a comparison with the bound in Eq.~(\ref{eq:tripletbound}). Suppose the SM Lagrangian is 
amended by the linear combination of operators in Eq.~(\ref{eq:eliminateT})
with a coefficient $\frac{1}{\Lambda^2}$. Then the 90\% confidence level bound
is only $\Lambda>600~{\rm GeV} $, which we obtained using Ref.~\cite{Han:2004az}.
Clearly, we can accommodate a much larger contribution from the triplet scalar provided that
we generate the appropriate combination of operators.

Our goal in this toy example is obtaining the linear combination of operators in Eq.~(\ref{eq:eliminateT})
by including additional heavy states that induce the operators $O_{hf}^Y$, $O_{hl}^t$, and $O_{hq}^t$.
The obvious choices are heavy gauge bosons of additional spontaneously broken $SU(2) \times U(1)$
symmetry. We will refer to such gauge bosons as $B'$ and $W'$. Suppose that these bosons couple
to the Higgs current and the fermion currents as follows:
\begin{equation}
  {\mathcal L}=   q_h B'_\mu \, j^\mu_h + q_f  \sum_f Y_f  B'_\mu\,  j^\mu_f + q_W {W'}_\mu^a\,  j^{a\mu}_h +  
  q_W {W'}^a_\mu \sum_{f_L} \, j^{a \mu}_f,
\end{equation}
where $j^\mu_h = i ( h^\dagger D_\mu h - D_\mu h^\dagger h)$, 
$j^{a \mu}_h = i ( h^\dagger \sigma^a D_\mu h - D_\mu h^\dagger \sigma^a h)$,
and $j_f^\mu$ are the obvious fermion currents. We denoted the coupling constants and charges as
$q_{h,f,W}$ and assumed that the couplings of fermions to the $B'$ are proportional to the fermion hypercharge.
 Integrating out the heavy vector boson yields
\begin{equation}
  {\mathcal L}_{B'W'}= - \frac{2 q_h^2}{M^2_{B'}} O_h - \frac{q_h q_f}{M^2_{B'}} O_{hf}^Y -
              \frac{q_W^2}{M_{W'}^2} (O_{hl}^t+ O_{hq}^t) + O_{4-f},
\end{equation}
where $O_{4-f}$ are four-fermion operators induced by the $B'$ and $W'$ that we do not need
to specify in detail.

It is clear that by choosing the couplings $q_{h,f,W}$ appropriately the sum of ${\mathcal L}_\phi$
and ${\mathcal L}_{B'W'}$  can be made proportional to the combination of operators on the left-hand side of  Eq.~(\ref{eq:eliminateT}). This would result in a theory with apparent custodial symmetry
breaking that is nevertheless relatively poorly  constrained. The only observable consequence
of the custodial symmetry breaking would be through the presence of the operator $O_{3V}$
on the right-hand side of Eq.~(\ref{eq:eliminateT}). This is a toy example because
of two obvious caveats. First, we had to fine-tune the couplings to obtain the desired linear combination
of different operators. To make it useful one would hope for a dynamical reason for the couplings and
masses of heavy fields to have suitable values. Second, the exchanges of the $B'$ and $W'$ also
induce four-fermi operators. Such operators are usually tightly constrained as well~\cite{Han:2004az}.
Of course, our toy example is reminiscent of the littlest Higgs model~\cite{Arkani-Hamed:2002qy},
where the heavy fields are analogous to the ones in our toy example. The details of the
effective operators induced in that case and bounds on the parameters are
presented in Ref.~\cite{Han:2005dz}.

One may expect that it is possible to move beyond toy models in the context of extra dimensions
with fermions and gauge fields living in the bulk. In that case, depending on the bulk and boundary
couplings as well as the background geometry, one can manipulate the wave functions of different fields to
minimize the constraints on such models. Indeed a subset of our results was used in the literature,
for example in Refs.~\cite{Carena:2003fx,Agashe:2003zs,Cacciapaglia:2004jz}.

In Refs.~\cite{Carena:2003fx,Agashe:2003zs} models with the linearly realized electroweak
symmetry are considered. A field redefinition is used to shift the oblique parameters in presence 
of additional operators. The particular field redefinition used there is equivalent to a linear combination 
of Eqs.~(\ref{eq:fred1}) and (\ref{eq:fred2}) where these equations are added together with their relative weights 
proportional to the coupling constants $g'$ and $g$, respectively.

In the case of  Higgsless models where electroweak symmetry is nonlinearly realized
it was shown in Ref.~\cite{Cacciapaglia:2004jz} how to reduce the oblique parameters
by an equivalent linear combination that can be obtained from our 
Eqs.~(\ref{eq:nf1}) to (\ref{eq:nf3}). It was also pointed out in Ref.~\cite{Chivukula:2005ji} that 
reducing the oblique parameters leaves an imprint on triple gauge boson couplings, which is 
in complete agreement with our results. It is worth remembering, however,  that there is more
than one operator relation that can be used to lessen the effect of oblique parameters
and exploiting that fact could lead to construction of more successful models.

\section{Conclusions}
\label{sec:Conclusions}

We have explored operator relations derived from equations of motions in effective theories
of physics beyond the Standard Model. The resulting operator identities relate the oblique parameters
 and operators that change gauge boson-fermion couplings to operators that modify gauge
 boson self couplings. When  electroweak symmetry is linearly realized there are
 two such relations involving the oblique parameters $S$ and $T$. In case of nonlinearly
 realized symmetry, there are three relations that involve  $S$, $T$, and $U$. 

These particular combinations of operators can only be observed by measuring the triple gauge 
boson couplings. If one constructed an extension of the SM in which only these special combination
of operators are present, there would be no other way to distinguish this model from the SM
by making precision measurements. This presents an interesting opportunity and challenge 
for model building. It also means that a better measurement of the gauge boson self couplings
could be very useful for constraining new physics. 

To date, some of the operator relations have been used to reduce electroweak constraints
on extra dimensional models. One accomplishes that by choosing profiles of fields to 
alter the couplings of the SM fields to Kaluza-Klein excitations. Not all operators relations
we pointed out  seem to have been explored in model building thus far. It would be intriguing if
one could find theories with dynamics such that only the special linear combinations of
operators appear at low energies.

\section*{Acknowledgments}
WS thanks the members of SPhT at Saclay for their hospitality. We thank Zhenyu Han
for discussions about the effective theory with linearly realized electroweak symmetry, 
Thomas Hambye for discussions about nonlinear realizations,
 and Riccardo Rattazzi and Marco Serone for discussions on oblique corrections. 
 The research of CG was supported in part by the RTN European Program MRTN-CT-2004-503369
 and by the ACI Jeunes Chercheurs 2068, while that of WS was supported in part by the US Department of Energy
under grant  DE-FG02-92ER-40704,  and JT was supported by the US Department of Energy
under grant  DE-FG02-91ER40674.

\appendix
 \section{Electroweak chiral Lagrangian}
 \label{app:Lagrangian}
For completeness, we list here CP-conserving operators in the electroweak
chiral Lagrangian to the lowest interesting order. These include dimension 2 and 4 
operators containing the $\Sigma$ field and the gauge field strengths, 
as well as dimension 4 operators containing two fermions and
the $\Sigma$ field~\cite{Longhitano,AppelquistWu,ABCH,Bagan:1998vu}.
We briefly explain how to obtain these results.

It is useful to express the operators in terms of $V_\mu \equiv (D_\mu \Sigma) \Sigma^\dagger$
and $T\equiv \Sigma \sigma^3 \Sigma^\dagger$ instead of the $\Sigma$ field itself. 
We will also use  $\hat{V}_\mu \equiv (D_\mu \Sigma^\dagger) \Sigma$ and $\hat{T}\equiv \sigma^3$
when we discuss operators containing fermions. Under the $SU(2)_L \times SU(2)_R$
these combinations transform as follows
\begin{displaymath}
   (V_\mu,T) \longrightarrow L (V_\mu,T) L^\dagger \ \ {\rm and} \ \ 
   (\hat{V}_\mu,\hat{T}) \longrightarrow R (\hat{V}_\mu,\hat{T}) R^\dagger .
\end{displaymath}
Since $\Sigma$ is unitary $(D_\mu \Sigma) \Sigma^\dagger + \Sigma (D_\mu \Sigma^\dagger)=0$.
Therefore, $V_\mu^\dagger=-V_\mu$ and $T^\dagger=T$. The same holds for $\hat{V}_\mu$
and $\hat{T}$.  Because $\det(\Sigma)=1$ we have $\tr(V_\mu)=\tr(\hat{V}_\mu)=0$. $T$ and $\hat{T}$
are traceless as well.

Gauge invariant operators containing only the $\Sigma$ field and the gauge fields
can be written as traces of $V_\mu$, $T$, $B_{\mu\nu}$, and 
$W_{\mu \nu} =W^a_{\mu \nu} \frac{\sigma^a}{2}$. One could also construct invariants
by taking determinants instead, but doing so does not lead to any independent 
operators~\cite{Longhitano}. There is no need to use $\hat{V}_\mu$ and $\hat{T}$ since
traces are cyclic and $\Sigma$ is unitary. Also, there is no need to include covariant
derivatives because  derivatives acting on $B_{\mu\nu}$ and $W_{\mu \nu}$ can be
 removed by integration by parts. Moreover, $D_\mu T=[V_\mu,T]$ and
$D_\mu V_\nu - D_\nu V_\mu = - i g W_{\mu \nu} + \frac{i}{2} g' B_{\mu \nu} T + [V_\mu,V_\nu]$.
Our sign convention for the covariant derivative is such that $D_\mu \Sigma =
\partial_\mu \Sigma - i g W_\mu \Sigma + \frac{i}{2} g' B_\mu \Sigma \sigma^3$.   In addition, 
$\partial^\mu \tr(V_\nu T)\approx 0$ and $D^\mu V_\mu \approx 0$ by taking derivatives
of the equations of motion in Eq.~(\ref{eq:nBeom}) and (\ref{eq:nWeom}). 
The approximate sign indicates that we neglect terms proportional to fermion masses. Since
$W_{\mu \nu}$, $V_\mu$ ,and $T$ are traceless two-by-two matrices therefore an arbitrary trace of these
matrices can be written as a product of traces of pairs of matrices if  the number of matrices in such a trace
is even or as a product of one trace of three matrices and several traces of pairs of matrices otherwise.

Using the observations outlined above it is straightforward to enumerate operators of dimension
2:
\begin{eqnarray}
  {\mathcal L}_{kin}&=&-\frac{v^2}{4}\tr(V_\mu V^\mu), \\
  {\mathcal L}_0 &=&\frac{v^2}{4} \tr(V_\mu T) \tr(V^\mu T),
\end{eqnarray}
and dimension 4
\begin{eqnarray}
  {\mathcal L}_1 &=& \frac{1}{2} g g' B^{\mu \nu} \tr(W_{\mu \nu} T),  \\
  {\mathcal L}_2 &=&  \frac{i}{2} g' B^{\mu \nu} \tr([V_\mu,V_\nu]T),  \\
  {\mathcal L}_3 &=& i  g \tr(W^{\mu \nu} [V_\mu,V_\nu]),  \\
  {\mathcal L}_4 &=& \left[\tr(V_\mu V_\nu)\right]^2,  \\
  {\mathcal L}_5 &=&  \left[\tr(V^\mu V_\mu)\right]^2, \\
  {\mathcal L}_6 &=& \tr(V^\mu V^\nu) \tr(V_\mu T)  \tr(V_\nu T),  \\
  {\mathcal L}_7 &=&  \tr(V^\mu V_\mu) \tr(V^\nu T)  \tr(V_\nu T),  \\
  {\mathcal L}_8 &=&  \frac{1}{4} g^2 \left[\tr(W_{\mu\nu} T)\right]^2,  \\
  {\mathcal L}_9 &=&  \frac{i}{2}  g \tr(W^{\mu\nu}T) \tr([V_\mu,V_\nu] T),   \\
  {\mathcal L}_{10} &=&  \frac{1}{2}  \left[\tr(V^\mu T) \tr(V_\mu T )  \right]^2, \\
  {\mathcal L}_{11} &=&  g \epsilon^{\mu \nu \rho \sigma} \tr(V_\mu T) \tr(V_\nu W_{\rho \sigma}).
\end{eqnarray}
We followed the notation of Ref.~\cite{AppelquistWu}. Note that $i \tr(W^{\mu \nu} [V_\mu, T]) \tr(V_\nu T)$ 
can be expressed as a combination of $ {\mathcal L}_3$ and $ {\mathcal L}_9$.
All operators in the equations above are Hermitian and even under CP. Under CP,
the fields of interest to us have the following transformation properties
\begin{eqnarray}
  (CP) \, B_{\mu \nu} \, (CP)^{-1} = - (-1)^{\mu+\nu} B_{\mu\nu} 
         &\ \ & (CP) \, W_{\mu \nu} \, (CP)^{-1} =  (-1)^{\mu+\nu} \sigma^2 W_{\mu \nu} \sigma^2 \\
    (CP) \, V_\mu \, (CP)^{-1} = (-1)^\mu \sigma^2 V_\mu \sigma^2
          &\ \ & (CP) \, T \, (CP)^{-1}=-\sigma^2 T \sigma^2, \label{eq:CP}
 \end{eqnarray}
 where the last line follows from $(CP) \, \pi^a \sigma^a \, (CP)^{-1}=\sigma^2 (\pi^a \sigma^a)\sigma^2 $.
 Meanwhile, $(-1)^\mu$ equals $1$ for $\mu$ in the time direction and $-1$  for $\mu$ in the space directions.
 Therefore, terms without the $\epsilon^{\alpha \beta \gamma \delta}$ tensor are even under CP
 when the combined number of $B_{\mu \nu}$ and $T$ fields is even. Terms with the 
 $\epsilon^{\alpha \beta \gamma \delta}$ tensor must have an odd number of $B_{\mu \nu}$ and $T$ fields 
 to be CP even.
 
 We now turn to operators of dimension 4 containing two fermions~\cite{ABCH,Bagan:1998vu}.
 These are all products of fermion currents with one $V_\mu$, or $\hat{V}_\mu$, 
 and several $T$'s, or $\hat{T}$'s, sandwiched between the fermions. There are six such operators
 \begin{eqnarray}
   {\mathcal L}^1_f &=&i \overline{f}_L \gamma^\mu V_\mu f_L, \\
   {\mathcal L}^2_f &=&i \overline{f}_L \gamma^\mu (V_\mu T + T V_\mu) f_L, \\
   {\mathcal L}^3_f &=&i \overline{f}_L \gamma^\mu  T V_\mu T f_L, \\
   {\mathcal L}^4_f &=&i \overline{f}_R \gamma^\mu \hat{V}_\mu f_R, \\
   {\mathcal L}^5_f &=&i \overline{f}_R \gamma^\mu ( \hat{V}_\mu \hat{T} + \hat{T} \hat{V}_\mu) f_R, \\
   {\mathcal L}^6_f &=&i \overline{f}_R \gamma^\mu \hat{T} \hat{V}_\mu \hat{T} f_R.
\end{eqnarray}
The subscript $f$ can be either $q$ or $l$ for operators ${\mathcal L}^{1,2,3}$.  With a slight abuse
of notation, we are going to use the same subscript for ${\mathcal L}^{4,5,6}$ to denote the doublets of 
right-handed quarks or leptons.

Operators of the form fermion current times a trace of the bosonic matrices are not independent.
If the current and the trace contain a sum over $\sigma^a$ such operator can be reduced to
${\mathcal L}^{1\ldots 6}_f$ using the completeness relation for the Pauli matrices.
Without the Pauli matrices there is only one  non-vanishing trace that is $\tr(V_\mu T)$. 
Because $V_\mu$ and $T$ are traceless $V_\mu T+T V_\mu$ is proportional to the identity matrix, 
so its trace is trivial. The operators with fermions, 
${\mathcal L}^{1\ldots 6}_f$, are CP even. The standard CP transformation property of fermion currents,
 $(CP) \, j^\mu_{L,R} \, (CP)^{-1} = - (-1)^\mu j^\mu_{L,R}$, involves transposing  the fermion fields.
Transposing the right-hand sides of Eq.~(\ref{eq:CP}) yields extra minus signs:
$(\sigma^2 V_\mu \sigma^2)^T = - V_\mu$ and $(\sigma^2 T \sigma^2)^T =-T$. The CP transformation
properties of $\hat{V}_\mu$ and $\hat{T}$ are identical to those of $V_\mu$ and $T$.



\begin{thebibliography}{99}


\bibitem{oblique}
M.~Golden and L.~Randall,
Nucl.\ Phys.\ B {\bf 361}, 3 (1991);
B.~Holdom and J.~Terning,
Phys.\ Lett.\ B {\bf 247}, 88 (1990);
M.~E.~Peskin and T.~Takeuchi,
Phys.\ Rev.\ Lett.\  {\bf 65}, 964 (1990);
G.~Altarelli and R.~Barbieri,
Phys.\ Lett.\ B {\bf 253}, 161 (1991).

\bibitem{STU}
M.~E.~Peskin and T.~Takeuchi,
Phys.\ Rev.\ D {\bf 46}, 381 (1992).

\bibitem{LEP12}
R.~Barbieri, A.~Pomarol, R.~Rattazzi and A.~Strumia,
Nucl.\ Phys.\ B {\bf 703}, 127 (2004)
[arXiv:hep-ph/0405040].

\bibitem{Han:2004az}
Z.~Han and W.~Skiba,
Phys.\ Rev.\ D {\bf 71}, 075009 (2005)
[arXiv:hep-ph/0412166].

\bibitem{Previous}
 G.~Sanchez-Colon and J.~Wudka,
  Phys.\ Lett.\ B {\bf 432}, 383 (1998)
  [arXiv:hep-ph/9805366];
%
W.~Kilian and J.~Reuter,
  Phys.\ Rev.\ D {\bf 70}, 015004 (2004)
  [arXiv:hep-ph/0311095].

\bibitem{Buchmuller}
W.~Buchmuller and D.~Wyler,
Nucl.\ Phys.\ B {\bf 268}, 621 (1986).

\bibitem{Hagiwara:1986vm}
K.~Hagiwara, R.~D.~Peccei, D.~Zeppenfeld and K.~Hikasa,
Nucl.\ Phys.\ B {\bf 282}, 253 (1987).

\bibitem{field:red}
H.~D.~Politzer,
Nucl.\ Phys.\ B {\bf 172}, 349 (1980);
H.~Georgi,
Nucl.\ Phys.\ B {\bf 361}, 339 (1991).

\bibitem{Strumia:1999jm}
  A.~Strumia,
  Phys.\ Lett.\ B {\bf 466}, 107 (1999)
  [arXiv:hep-ph/9906266].
  
\bibitem{AppelquistBernard}
T.~Appelquist and C.~W.~Bernard,
Phys.\ Rev.\ D {\bf 22}, 200 (1980).

\bibitem{Longhitano}
A.~C.~Longhitano,
Phys.\ Rev.\ D {\bf 22}, 1166 (1980);
Nucl.\ Phys.\ B {\bf 188}, 118 (1981).

\bibitem{AppelquistWu}
T.~Appelquist and G.~H.~Wu,
Phys.\ Rev.\ D {\bf 48}, 3235 (1993)
[arXiv:hep-ph/9304240].

\bibitem{ABCH}
T.~Appelquist, M.~J.~Bowick, E.~Cohler and A.~I.~Hauser,
Phys.\ Rev.\ D {\bf 31}, 1676 (1985).

\bibitem{Bagan:1998vu}
E.~Bagan, D.~Espriu and J.~Manzano,
Phys.\ Rev.\ D {\bf 60}, 114035 (1999)
[arXiv:hep-ph/9809237].

\bibitem{Nyffeler:1999ap}
A.~Nyffeler and A.~Schenk,
Phys.\ Rev.\ D {\bf 62}, 113006 (2000)
[arXiv:hep-ph/9907294].

\bibitem{Arkani-Hamed:2002qy}
N.~Arkani-Hamed, A.~G.~Cohen, E.~Katz and A.~E.~Nelson,
JHEP {\bf 0207}, 034 (2002)
[arXiv:hep-ph/0206021].

\bibitem{Han:2005dz}
Z.~Han and W.~Skiba,
Phys.\ Rev.\ D {\bf 72}, 035005 (2005)
[arXiv:hep-ph/0506206].

\bibitem{Carena:2003fx}
  M.~Carena, A.~Delgado, E.~Ponton, T.~M.~P.~Tait and C.~E.~M.~Wagner,
  Phys.\ Rev.\ D {\bf 68}, 035010 (2003)
  [arXiv:hep-ph/0305188].
  
\bibitem{Agashe:2003zs}
K.~Agashe, A.~Delgado, M.~J.~May and R.~Sundrum,
JHEP {\bf 0308}, 050 (2003)
[arXiv:hep-ph/0308036].

\bibitem{Cacciapaglia:2004jz}
G.~Cacciapaglia, C.~Csaki, C.~Grojean and J.~Terning,
  Phys.\ Rev.\ D {\bf 70}, 075014 (2004)
  [arXiv:hep-ph/0401160]
and
Phys.\ Rev.\ D {\bf 71}, 035015 (2005)
[arXiv:hep-ph/0409126].

\bibitem{Chivukula:2005ji}
  R.~S.~Chivukula, E.~H.~Simmons, H.~J.~He, M.~Kurachi and M.~Tanabashi,
  Phys.\ Rev.\ D {\bf 72}, 075012 (2005)
  [arXiv:hep-ph/0508147].



\end{thebibliography}
\end{document}